\documentclass[conference,a4paper]{IEEEtran}
\usepackage{cite}
\usepackage{float}
\usepackage{graphicx}
\graphicspath{{./images/}}
\usepackage[T1]{fontenc}
\usepackage{amsmath}
\usepackage{xcolor}
\usepackage{mathtools}
\usepackage{multicol}
\usepackage{array} 
\usepackage{balance}
\usepackage{xcolor}
\usepackage{siunitx, booktabs}
\usepackage{ragged2e}

\newcommand{\ts}{\textsuperscript}

\begin{document}

\twocolumn[
  \begin{@twocolumnfalse}
  ©  20xx  IEEE.  Personal  use  of  this  material  is permitted.Permission from IEEE must be obtained for all other uses, in any current or future media, including reprinting/republishing this material for advertising or promotional purposes, creating new collective works, for resale or redistribution to servers or lists,  or  reuse  of any  copyrighted  component  of  this  work  in other works.
  \end{@twocolumnfalse}
]

\title{Modified Auto Regressive Technique for Univariate Time Series Prediction of Solar Irradiance}

\author{\IEEEauthorblockN{Umar Marikkar, A. S. Jameel Hassan, Mihitha S. Maithripala\\ Roshan I. Godaliyadda, Parakrama B. Ekanayake, Janaka B. Ekanayake\\
}
\IEEEauthorblockA{\emph{Department of Electrical and Electronic Engineering, Faculty of Engineering}\\ \emph{University of Peradeniya, Peradeniya 20400, Sri Lanka}\\}
\IEEEauthorblockA{ \tt{\{e14219, jameel.hassan.2014, e14215\}@eng.pdn.ac.lk}\\
\tt{\{roshangodd, mpb.ekanayake, jbe\}@ee.pdn.ac.lk}
}
}

\maketitle

\begin{abstract}
The integration of renewable resources has increased in power generation as a means to reduce the fossil fuel usage and mitigate its adverse effects on the environment. However, renewables like solar energy are stochastic in nature due to its high dependency on weather patterns. This uncertainty vastly diminishes the benefit of solar panel integration and increases the operating costs due to larger energy reserve requirement. To address this issue, a Modified Auto Regressive model, a Convolutional Neural Network and a Long Short Term Memory neural network that can accurately predict the solar irradiance are proposed. The proposed techniques are compared against each other by means of multiple error metrics of validation. The Modified Auto Regressive model has a mean absolute percentage error of 14.2\%, 19.9\% and 22.4\% for 10 minute, 30 minute and 1 hour prediction horizons. Therefore, the Modified Auto Regressive model is proposed as the most robust method, assimilating the state of the art neural networks for the solar forecasting problem.\par


\end{abstract}

\section{Introduction}
Over the years, penetration of renewable energy has vastly increased in the electricity grid network. The increase in energy demand, adverse effects of fossil fuel generation and awareness towards climate change has advanced the use of renewable resources \cite{abbot}. The growing concern towards environmental pollution has rendered the sustainable development goal \#7 aiming to develop cleaner energy sources by the United Nations environment program \cite{UNDP}. \par

The increased intervention by renewable resources such as solar and wind pose a highly volatile problem due to the intermittent nature of power generation. In comparison, solar power poses a bigger problem than wind power due to higher fluctuations due to cloud covers and effects of instant weather changes. There is a critical need in improving the real time solar irradiance forecast  since the core operations of the utility and Independent System Operators (ISO) depends on the power generation capacity. For instance, the day ahead prediction of solar irradiance will make the Unit Commitment (UC) more efficient whereas the improvement in short term forecast will reduce errors due to fluctuations in solar power and minimise the strain on the grid. This indicates that a higher prediction accuracy is useful in many timescales.\par

The integration of renewable energy is inevitable in the ever growing energy demand and is a huge contribution to the smart grid evolution. For a viable progress of the smart grid, this integration needs to be addressed in terms of its stakeholders; the ISO and Independent Power Producers (IPP). In the energy market, the ISO can dispatch and share resources amongst IPPs for the energy bidding process. In order to optimally utilise the power production, the IPPs need to minimise their error of power production forecast. A lower forecast error, gives the IPP a larger window of bidding during the intra-day energy bidding process in light of the solar power prediction. This intra-day bidding ranges in window time from 15 minutes to a couple of hours. However, failure to produce the bid power will be settled in terms of cumbersome penalties for the energy bidder \cite{kaur}. Furthermore, these predictions are seldom deployed in-house which provide the specific forecast horizon depending on the necessity of the power producer. The critical feature for short-term prediction schemes is the time constraint in data acquisition and processing \cite{energybid}. Having a very narrow window needs to ensure that the error is minimised by a very quick operating forecast technique. Thus, the robustness of the forecast scheme will depend not only on its accuracy of prediction but also on the ease of deployability having an instant processing capability.\par

The solar prediction scheme can be categorised into long term, short-term and very short-term predictions. In the literature a day ahead solar irradiance prediction has  been  performed  using  neural  networks  considering  the ground  sensor  data  and  weather  information  as  input in \cite{lasso_lstm}.  A combined  neural  network  approach  using  Convoluted  Neural Networks (CNN)  and  Long  Short  Term  Memory (LSTM)  is used  for  day  ahead  prediction  in \cite{woonghee}. A day ahead probabilistic Photo Voltaic (PV) power forecasting method based on auantile CNN based on feature extraction is discussed in \cite{huang}. A novel CNN framework using genetic algorithm and particle swarm optimisation for the hyper parameter tuning is presented in \cite{Nadong} using multiple meteorological data. Meanwhile, very short-term prediction ranging from few minutes to 6 hours is performed using multiple techniques including deep learning techniques, sky image based methods and statistical methods for time series data such as Auto Regression (AR), Auto Regression and Moving Average (ARMA) and Auto Regression Integrated Moving Average (ARIMA) \cite{lasanthika,skyimage,ar}. Statistical methods incorporating AR, ARIMA, ARMA have been extensively exercised in time series data in various other applications in literature \cite{Rafi,Matsila,Ningchen}. The statistical models AR and ARIMA are compared to LSTM in \cite{Ji}. The LSTM model is reported to outperform the statistical methods here. A hybrid model incorporating a discrete wavelet transform, ARMA and a Recurrent Neural Network (RNN) is implemented for 1 minute ahead prediction in \cite{nazaripouya} showing considerable improvement in precision of prediction.\par

In the literature, very short-term prediction of solar irradiance using deep learning techniques is consolidated to outperform others; specifically the LSTM neural network. However, a simplistic statistical forecast technique will be more robust in deployment due to the time constraint in very short-term prediction requirements. Therefore, in comparison to the deep learning techniques the statistical methods are convenient in computational expense. Furthermore, considering the above contributions, the usage of multiple data sources extensively is a major drawback in the ease of in-house implementation of the forecast schemes due to the scarcity of resources. In this paper, we propose a Modified Auto Regressive (MAR) approach, a CNN model and LSTM model for univariate solar prediction. The three models are compared across multiple error metrics for validation. Finally we propose the MAR as the best approach as it is able to assimilate its performance to the LSTM model for multiple prediction horizons as verified in this paper for 10 minute, 30 minute and 1 hour horizons using only the past irradiance measurements as inputs. This ensures a highly robust model that is easily deployable in-house, for real time very short-term predictions.  

\section{Data Preparation}
\label{sec:data}
\subsection{Study Area}
The solar irradiance data was obtained from the PV plant stationed at the Faculty of Engineering in University of Peradeniya in Sri Lanka. The city of Peradeniya is located surrounded by the hills of Hantana with a tropical climate. This results in fluctuations of the solar irradiance curve rather than yielding the typical \emph{"bell"} shaped curve. This setting gives a more challenging data set which highly reflects the volatile nature of solar irradiance in contrast to data sets often encountered in the literature. The data is collected for a period of one year with data points at every 10 minute interval. \par

\subsection{Training/Testing Split and Data Standardisation}
\label{subsection:preprocess}
For all forecasting models, the training/testing data split is divided as 70/30$\%$ considering conventional deep learning practice. As the collected data spans a whole year, this gives a sufficiently large data-set ($\approx$110 days) for testing.\par
For an efficient training and forecast performance, the input data is standardised as in equation (\ref{eq:standardize}) as a pre processing step, and de-standardised in the post processing stage.\par
\begin{equation}
    \label{eq:standardize}
    z=\frac{x-\mu}{\sigma}
\end{equation}
where, \newline
\begin{tabular}{ll}
   $z$ &= Normalised signal value\\
   $x$ &=   Irradiance level at each timestamp\\
   $\mu$ &= Mean of the dataset\\
   $\sigma$ &=  Variance of the dataset\\
\end{tabular}

\section{Methodology}
\label{sec:method}
The short-term prediction of solar irradiance is implemented for the time horizons of 10 minute, 30 minute and 1 hour intervals. The forecasting schemes are developed using machine learning techniques in terms of Convolutional Neural Networks (CNN), Long-Short Term Memory (LSTM) networks, and in addition a Modified Auto Regressive (MAR) forecast model is implemented. Out of the three techniques, the MAR approach is highlighted as the best model for solar prediction.\par

\subsection{Convolutional Neural Network (CNN)}
\label{subsec:CNN}
CNNs are a type of neural networks most prominently deployed in image processing and classification problems. The key notion of CNN is its ability to learn abstract features and recognise them during training by means of kernels in the network \cite{oehmcke}. Therefore, in this paper, the CNN has been employed in a time series model to identify the temporal abstract level features in order to predict the next time step. \par

In order to encapsulate the complex modelling of features, the CNN utilises three separate layers, namely: the convolutional layer, the pooling layer and the fully connected layer. The convolutional layer is responsible for the identification of relationships between the inputs in the locality of the convolution operation that takes place between inputs and the kernels. The pooling layer performs a down-sampling of the output from the convolution operation. This is then fed to a fully connected layer which is responsible for predicting the output depending on the features. A series of convolution-pooling layers can be used if necessary.\par

In this paper, one convolution layer and an average pooling layer is used. These layers are designed to extract the feature on the input variables, which is the past 4 samples (selected as in section \ref{subsubsec:AROrder}) of the time series sequence, as in equation (\ref{eq:CNN}).\par
\begin{equation}
    \label{eq:CNN}
    h^{k}_{ij} = g((W^{k}*x)_{ij} + b_k)\\
\end{equation}
where, $W^{k}$ is the weight of the kernel connected to $k^{th}$ feature map, $g$ is the activation function and $b_k$ is the bias unit. \par

The Rectified Linear Unit (ReLU) function is used as the activation function after evaluating the performance against other activation functions. The ReLU function is defined by equation (\ref{eq:Relu}).
\begin{equation}
        \label{eq:Relu}
    g(x) = max(0,x)
\end{equation}

\begin{figure}[b]
    \centering
    \includegraphics[width=\columnwidth]{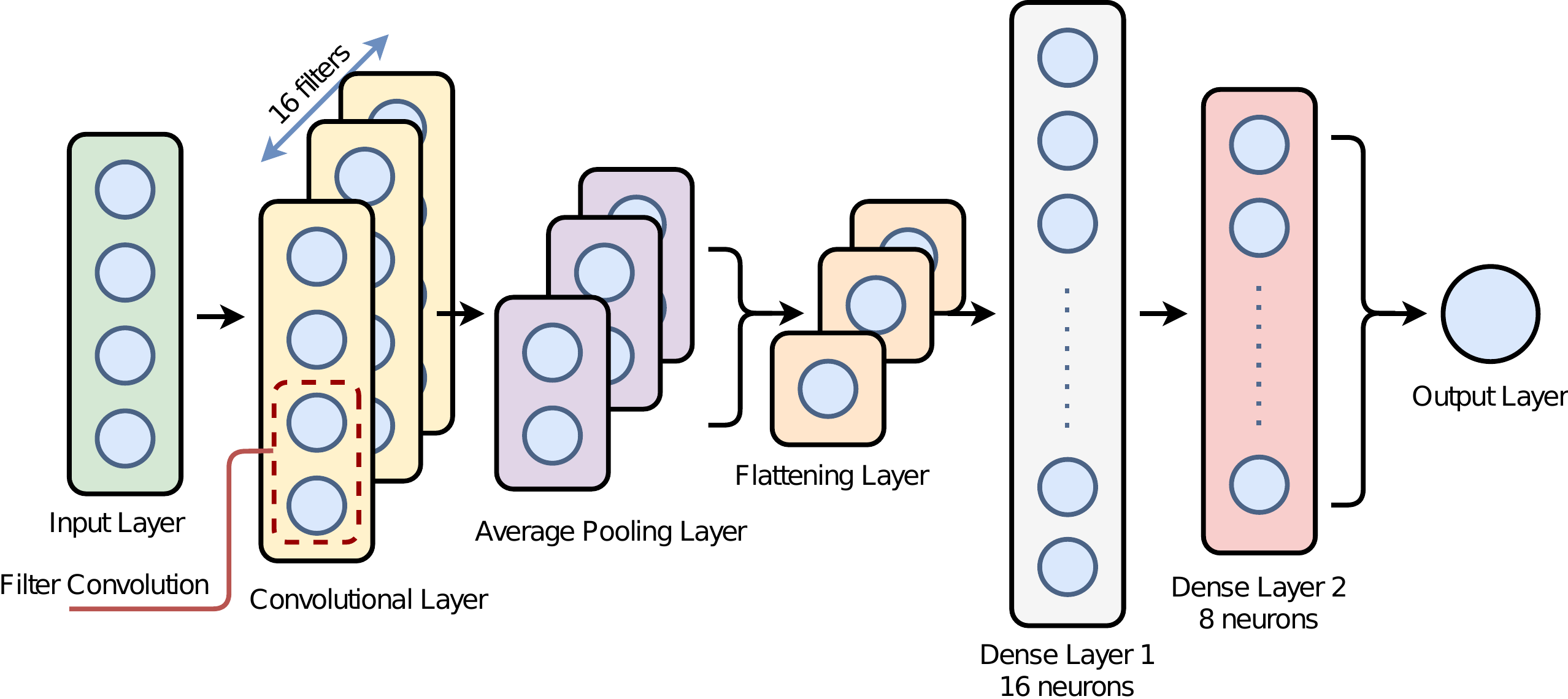}
    \caption{CNN architecture}
    \label{fig:cnn}
\end{figure}

The Adam optimisation algorithm is used as the training function which is an efficient implementation of the gradient descent algorithm \cite{Wen}. Finally, two dense - fully connected layers are implemented following the pooling layer with one final dense layer with a single neuron that outputs the prediction. An abstraction of the CNN architecture implemented is shown in Fig \ref{fig:cnn}. The hyper parameters of the model are chosen by optimisation of a grid search algorithm as highlighted in Table \ref{table:HPtuning}.\par

\subsubsection{Pre processing and Post processing stages} \hfill \break 
The solar irradiance curve has a trend of the $"bell"$ shape to it. In order to remove this trend in the input data, pre processing is performed at the input stage. In addition to the data standardisation described in \ref{subsection:preprocess}, a difference transform of lag 1 is performed to the input signal after standardisation. The transformed input is fed to the CNN and the predicted signal is obtained. The predicted signal is passed in a post processing stage to reconstruct the solar irradiance curve as predicted. The pre processing difference transform and post processing reconstruction equations are given in equation (\ref{eq:difftransf}).
\begin{equation}
    \label{eq:difftransf}
    \begin{split}
        X &= [x_0 , x_1, \dots, x_n]\\
        \Tilde{X} &= [(x_0-0), (x_1-x_0), \dots, (x_n-x_{n-1})]\\
        \Tilde{Y} &= [\Tilde{y_0}, \Tilde{y_1}, \dots, \Tilde{y_n}]\\
        Y &= [(\Tilde{y_0}+0), (\Tilde{y_1}+x_0), \dots, (\Tilde{y_n} +x_{n-1})\\
    \end{split}
\end{equation}
here,\\
\begin{tabular}{ll}
$X$ &= Normalised signal value\\
$\Tilde{X}$ &= Difference transformed input\\
$\Tilde{Y}$ &= Predicted signal\\
$Y$  &= Reconstructed predicted signal value\\
\end{tabular}

\subsection{Long-Short Term Memory Neural Network (LSTM)}
\label{subsec:LSTM}
\begin{figure}[t]
    \centering
    \includegraphics[width=\columnwidth]{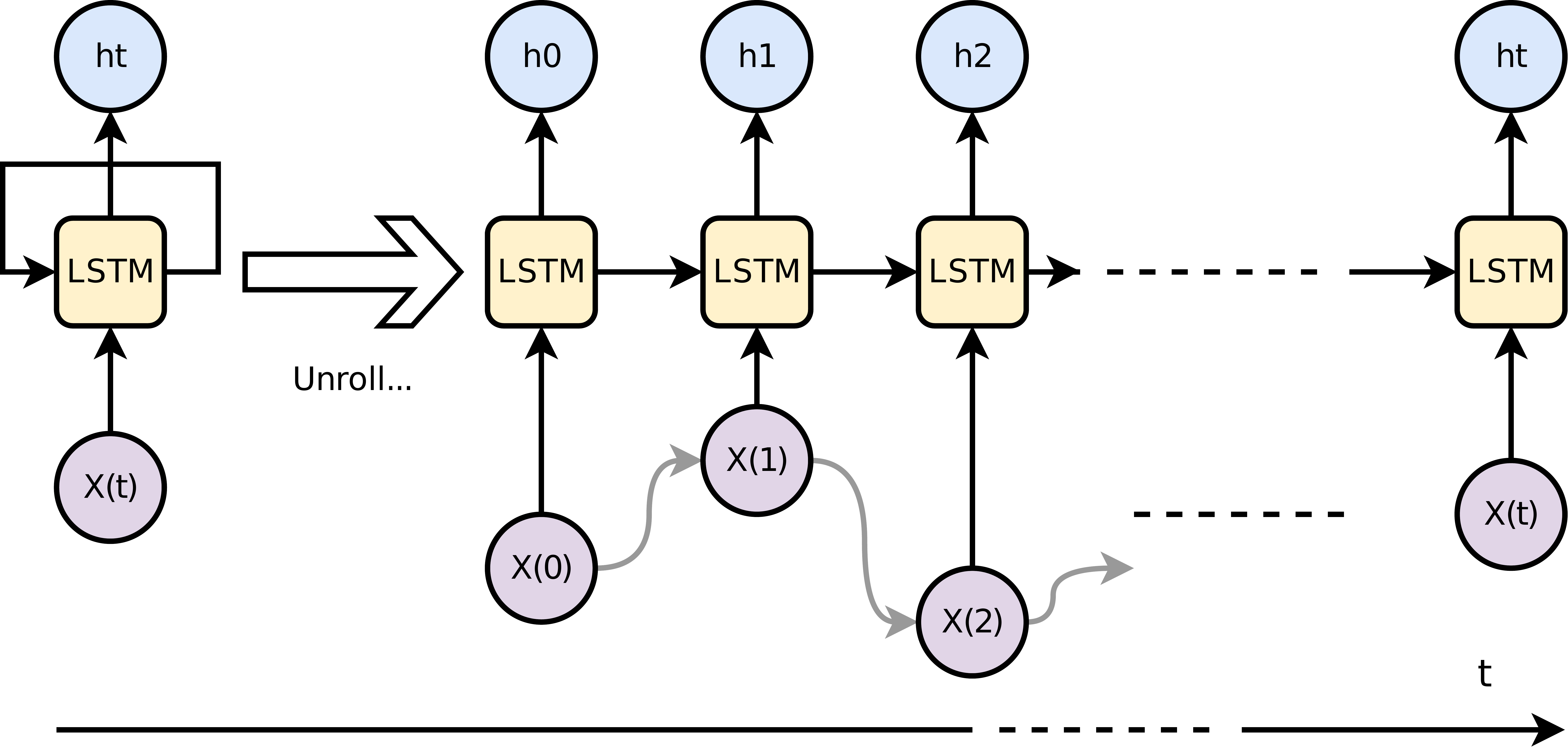}
    \caption{Working mechanism of LSTM cells}
    \label{fig:LSTM}
\end{figure}

\begin{figure}[b]
    \centering
    \includegraphics[width=0.7\columnwidth,scale=0.5]{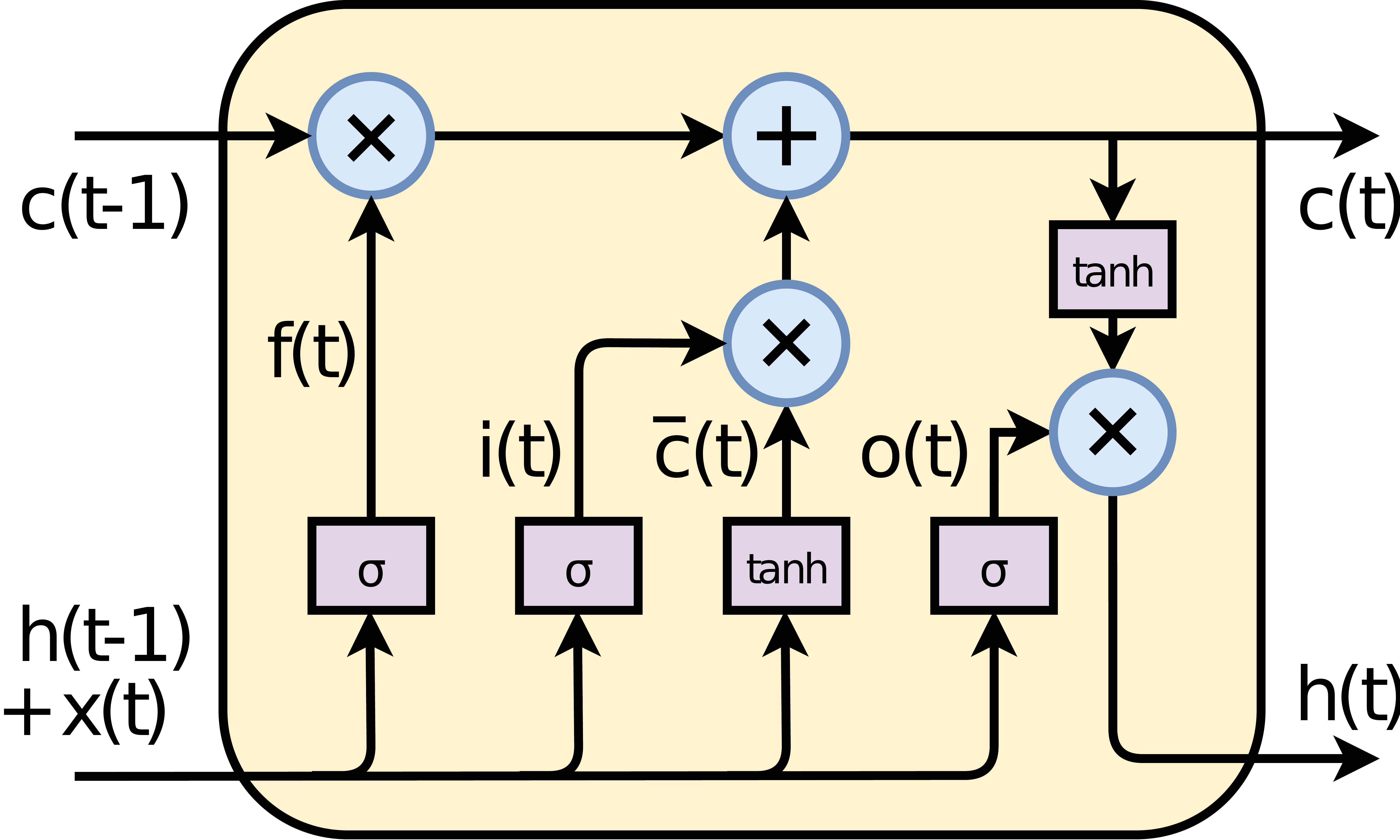}
    \caption{LSTM cell architecture}
    \label{fig:cell_LSTM}
\end{figure}

The LSTM network is a type of Recurrent Neural Networks (RNN), used for time series prediction. A major drawback of RNN is the inability to capture long-term dependencies in a signal, due to memory constraints. The LSTM cell has a selective storage of trends in its memory, hence it ignores repetitive information. The cell state is defined by which information is stored in or discarded. This is controlled by means of three gates; the input gate $i_t$, output gate $O_t$ and forget gate $f_t$. The output of the LSTM networks depends on the current input and the cell state \cite{hangx}. The working mechanism and cell architecture of the LSTM network is shown in Fig. \ref{fig:LSTM} and Fig. \ref{fig:cell_LSTM} respectively.\par

At time $t$, the inputs to the network are the sequence vector $X_{t}$, the hidden state output $h_{t-1}$ and the cell state $C_{t-1}$. The outputs of the network are, the LSTM hidden state $h_{t}$ and the cell state $C_{t}$. The forget gate, input gate and output gate are calculated using equations (\ref{eq:lstm1}), (\ref{eq:lstm2}) and (\ref{eq:lstm3}). Here, $i_{t}$ is the input gate and $O_{t}$ is the output gate. The forget gate $f_{t}$ is used to update, maintain or delete the cell state information. 
\begin{equation}
    \label{eq:lstm1}
    f_{t} = \sigma(W_{f}\times[h_{t-1},x_{t}]+b_{f})\\
\end{equation}
\begin{equation}
    \label{eq:lstm2}
    i_{t} = \sigma(W_{i}\times[h_{t-1},x_{t}]+b_{i})\\
\end{equation}
\begin{equation}
    \label{eq:lstm3}
    O_{t} = \sigma(W_{O}\times[h_{t-1},x_{t}]+b_{o})\\
\end{equation}
The current candidate cell state  $\bar{C}$ is calculated by equation (\ref{eq:lstm4}), and is updated to produce the output cell state $C_{t}$ as in equation (\ref{eq:lstm5}). Using the output cell state, the current hidden state $h_{t}$ is calculated by equation (\ref{eq:lstm6}).
\begin{equation}
    \label{eq:lstm4}
     \bar{C_{t}} = tanh(W_{C}\times[h_{t-1},x_{t}]+b_{c})\\
\end{equation}
\begin{equation}
    \label{eq:lstm5}
     C_{t} =f_{t}\times C_{t-1} + i_{t}\times \bar{C_{t}}\\
\end{equation}
\begin{equation}
    \label{eq:lstm6}
    h_{t} = O_{t}\times tanh(C_{t})\\
\end{equation}
 
$ W_{f} $, $ W_{i} $,  $ W_{O} $, and  $ b_{f} $, $ b_{i} $ , $ b_{o} $ and $b_{c}$ are weights and bias parameters of each gate. $\sigma$ is the sigmoid activation function.

\begin{table*}[t]
\caption{Hyper-parameter optimization for implemented networks}
\label{table:HPtuning}
\centering
\begin{tabular}{ m {0.2\linewidth} p{0.35\linewidth} p{0.35\linewidth}} 
    \toprule
    \textbf{Network Model}  & \textbf{Model Hyper-Parameter Names} & \textbf{Search Space for Optimal Hyper-Parameters}  \\
    \midrule
    CNN	            &Optimizer                  &Adam\\
                    &Learning rate ($\alpha$)	    &[0.1 0.01 \textbf{0.005} 0.001]   \\
                    &Convolution Layers	        &[\textbf{1} 2]   \\
                    &Fully connected Layers	     &[1 \textbf{2} 3]   \\
                    &Hidden Layer Neurons	    &[\textbf{8} \textbf{16} 50 150 300]  \\
                    &Number of Kernels	                &[3 \textbf{16} 80 150]   \\
                    &Kernel Size	            &[3 \textbf{2}]   \\
                    &Batch Size                 &[16 32 64 128 \textbf{256} 512]\\
                    &Pooling Size               &[\textbf{1} 2]\\
                    &Epochs                     &[10 \textbf{30} 100 500]\\
               
    \midrule
    LSTM	        &Optimizer                  &Adam\\
                    &Initial learning rate ($\alpha$) &[0.1 \textbf{0.05} 0.01 0.005]   \\
                    &Learning rate drop period  &[10 \textbf{30} 100 300]   \\
                    &LSTM Layers	            &[\textbf{1} 2]   \\
                    &LSTM Neurons	            &[8 16 \textbf{32} 64 128 256]  \\
                    &Fully Connected Layers	    &[\textbf{1} 2 3]   \\
                    &Hidden Layer Neurons	    &[\textbf{8} 16 32 64 128]  \\
                    &Epochs                     &[50 \textbf{100} 300 500]\\
    \bottomrule
\end{tabular}
\end{table*}

\subsection{Network Design for Deep-Learning Models}
\label{subsec:HPTuning}

All simulations are run on an Intel core-i7 @4.5GHz computer. Implemented deep-learning networks are designed using MATLAB deep learning toolbox.

Neural networks, if poorly trained, leads to over-fitting or under-fitting of the training data, resulting in disparity between training data prediction and actual prediction performance. Similarly, bad design of an neural network architecture could lead to error propagation, high computational cost, or simply overkill. \par
Hyper-parameter optimization plays an important role in choosing the optimal neural network architecture and training parameters. Brute force methods such as grid search, probabilistic models such as bayesian optimization and random searches are widely used. As high computational power is available for training, grid search algorithm was implemented. Initially, a coarse search was carried out on a large search space as shown in Table \ref{table:HPtuning}. Then, a fine search was implemented on a smaller search space. As all hyper-parameters were well optimized throughout the smaller search space, coarse search hyper-parameters were chosen, as highlighted in Table \ref{table:HPtuning}.

\subsection{Modified Auto-Regressive Model (MAR)}
\label{subsec:AR}

In the AR model, the predicted signal value at the next time step is linearly dependent on observed values at a set number of previous time steps. However, our proposed model does not work on the standardized irradiance measurements, but ensemble deducted values, as described in section \ref{subsubsec:features}. The AR model equation relating predicted value to the previously observed values is given by equation (\ref{eq:AR1}).
\begin{equation}
    \label{eq:AR1}
    x_{n,pred}=\sum_{k=1}^{m}w_{k} \times x_{n-k}
\end{equation}
where,\\
\begin{tabular}{ll}
  $m$   &= order of the AR model \\
  $x_{n,pred}$   &= predicted signal value for next timestamp\\
  $w_{k}$  &= model weights\\
  $x_{n-k}$  &= past signal values \\
  \\
\end{tabular}

\subsubsection{Feature Engineering by Ensemble Deduction} 
\begin{figure}[htb]
    \centering
    \includegraphics[width=\columnwidth]{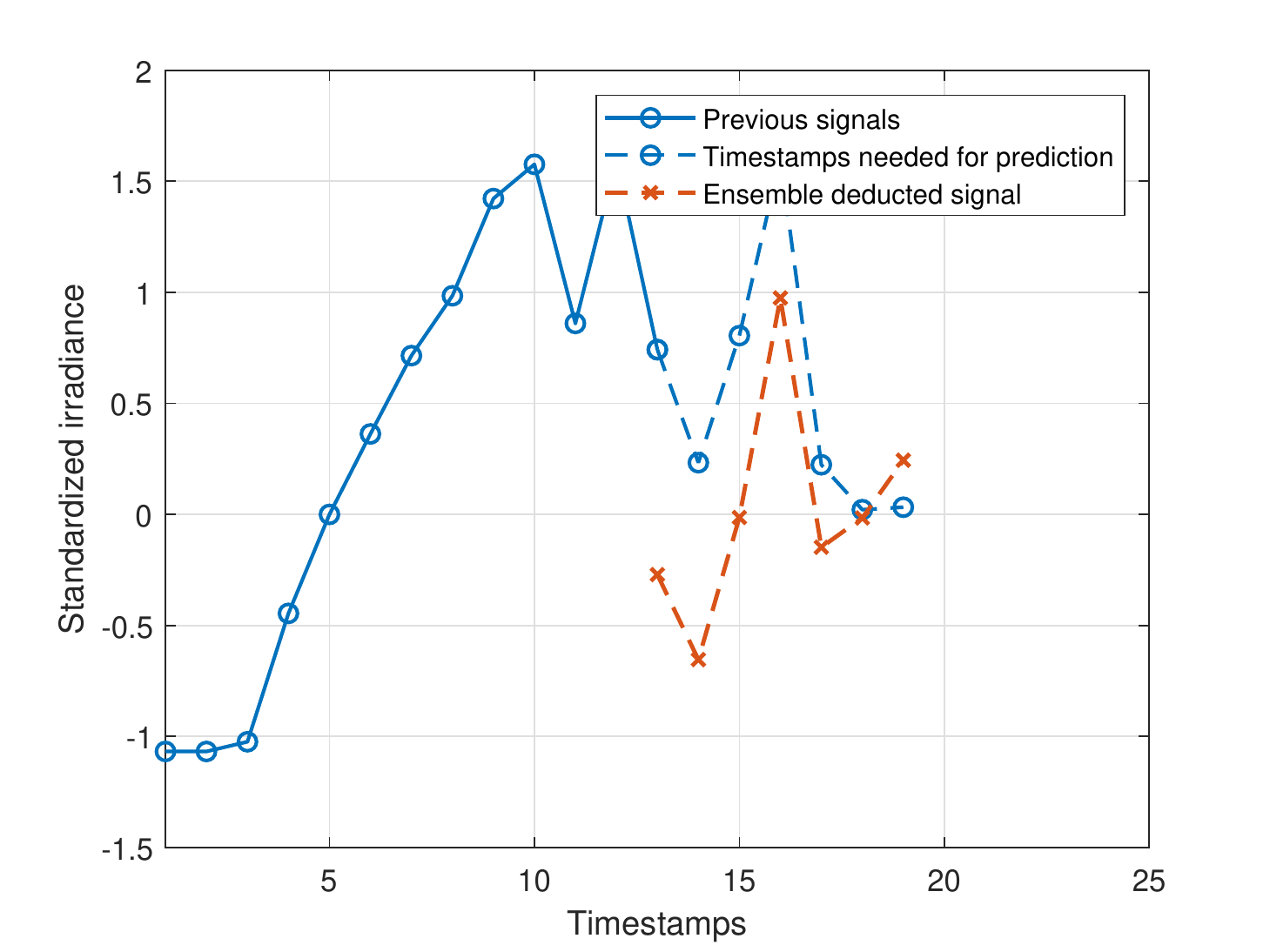}
    \caption{Ensemble Deduction to predict 20\ts{th} timestamp}
    \label{fig:ens}
\end{figure}

\hfill \break
\label{subsubsec:features}
Prior to the prediction, the expected value of $m$ number of past signal values at each timestamp is deducted from its corresponding irradiance measurement, as shown in equation (\ref{eq:AR2}). This ensures that the periodic nature of the days, governed by the bell shape curve, is unaffected at the time of prediction. The ensemble deduction in a given day to predict the 20\ts{th} timestamp of the day is illustrated in Fig. \ref{fig:ens}.   

\begin{equation}
    \label{eq:AR2}
    x_{n-i,ens} = x_{n-i} - E[x_{n-i}] \\
\end{equation}
where,\\
\begin{tabular}{ll}
  $i$   &= [1, \dots,$m$] \\
  $n$   &= prediction timestamp \\
  $x_{n-i,ens}$ &= ensemble deducted signal value at $n-i$\\
  $x_{n}$  &= actual standardized signal value at $n-i$ \\
  $E[.]$  &= statistical expectation operator \\
\end{tabular}
\newline

\subsubsection{Parameter Optimisation} 
\begin{figure}[htb]
    \centering
    \includegraphics[width=\columnwidth]{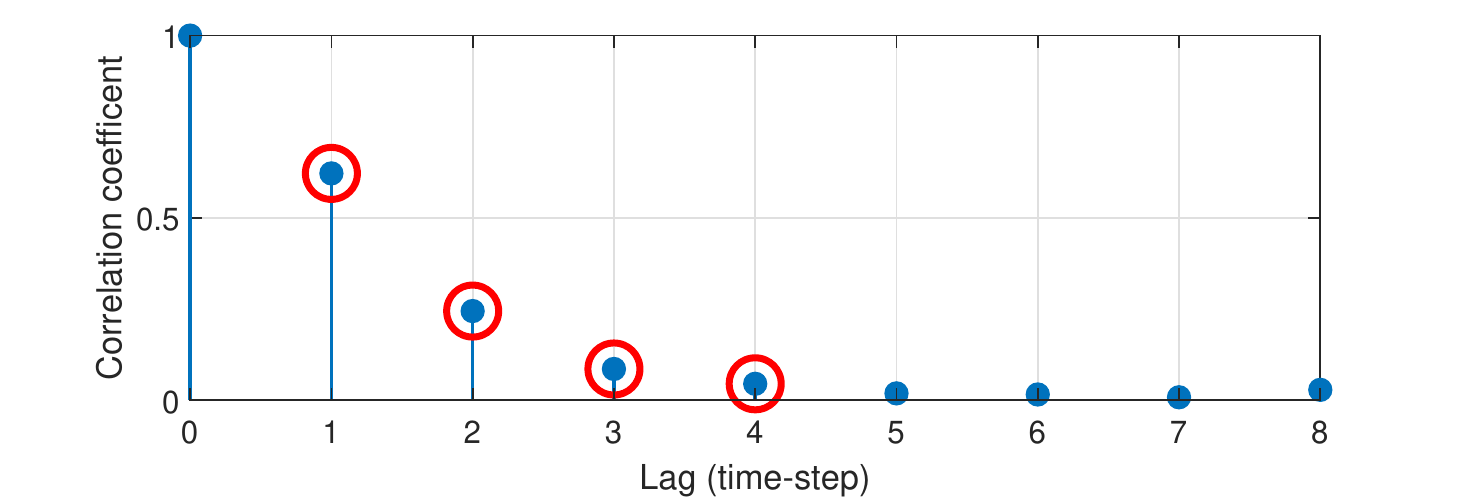}
    \caption{Partial Auto-Correlation Function (PACF)}
    \label{fig:PACF}
\end{figure}

\hfill \break
\label{subsubsec:AROrder}
The order of the model ($m$) depends on the Partial Auto-Correlation Function (PACF) of the given data. The PACF provides the correlation between a fixed time series value $x_{n}$ and its lagged values $x_{n-\tau}$ relative to the fixed value. The equation to compute the PACF is described in equation (\ref{eq:PACF}). Fig. \ref{fig:PACF} shows a graphical representation of equation (\ref{eq:PACF}). As observed, $m=4$ was chosen as the optimal order.   
\begin{equation}
    \label{eq:PACF}
    R_{\tau}=E[x_{n-\tau} \cdot x_{n}]
\end{equation}
where, \newline
\begin{tabular}{ll}
  $E[.]$   &= statistical expectation operator \\
  $R_{\tau}$ &= correlation function \\
  $\tau$   &= lag of the previous timestamp\\
\end{tabular}
\newline

The prediction error $x_{n,pred}-x_{n,real}$ is chosen to calculate model parameters. They are calculated using optimisation; where a positive, monotonically increasing error function is minimized. A squared error function as given by equation (\ref{eq:eq3}) exhibits these characteristics. Therefore, the Yule-Walker equation given by equation (\ref{eq:eq4}) is used to calculate model parameters.
\begin{equation}
    \label{eq:eq3}
    f(e_{n})= (x_{n,pred}-x_{n,real})^{2}
\end{equation}
where, \newline
\begin{tabular}{ll}
  $f(e_{n})$ &= error function \\
  $e_{n}$   &= error at a given time step $n$ \\
  $x_{n,pred}$   &= predicted value at $n$ \\
  $x_{n,real}$   &= observed value at $n$\\
\end{tabular}

\begin{equation}
    \label{eq:eq4}
    W=(X^{T}X)^{-1}X^{T}Y
\end{equation}
where,\newline
\begin{tabular}{ll}
    $W$ &= weights matrix \\
    $X$ &= design matrix (dependent on order $m$) \\
    $Y$ &= output matrix ($X_{real}$) \\
    \\
\end{tabular}

The design matrix $X$ contains the training examples as its rows, and features for each example as its columns. The number of columns depends on the order $m$. After optimizing the model parameters, a finite loop is run for each time step of the day, predicting the signal value $x_{n,pred}$ at the next time step. To calculate predicted solar irradiance, $x_{n,pred}$ is de-standardized.\par

\section{Results and Discussion}
\label{sec:results}

\begin{figure}[t]
    \centering
    \includegraphics[width=\columnwidth]{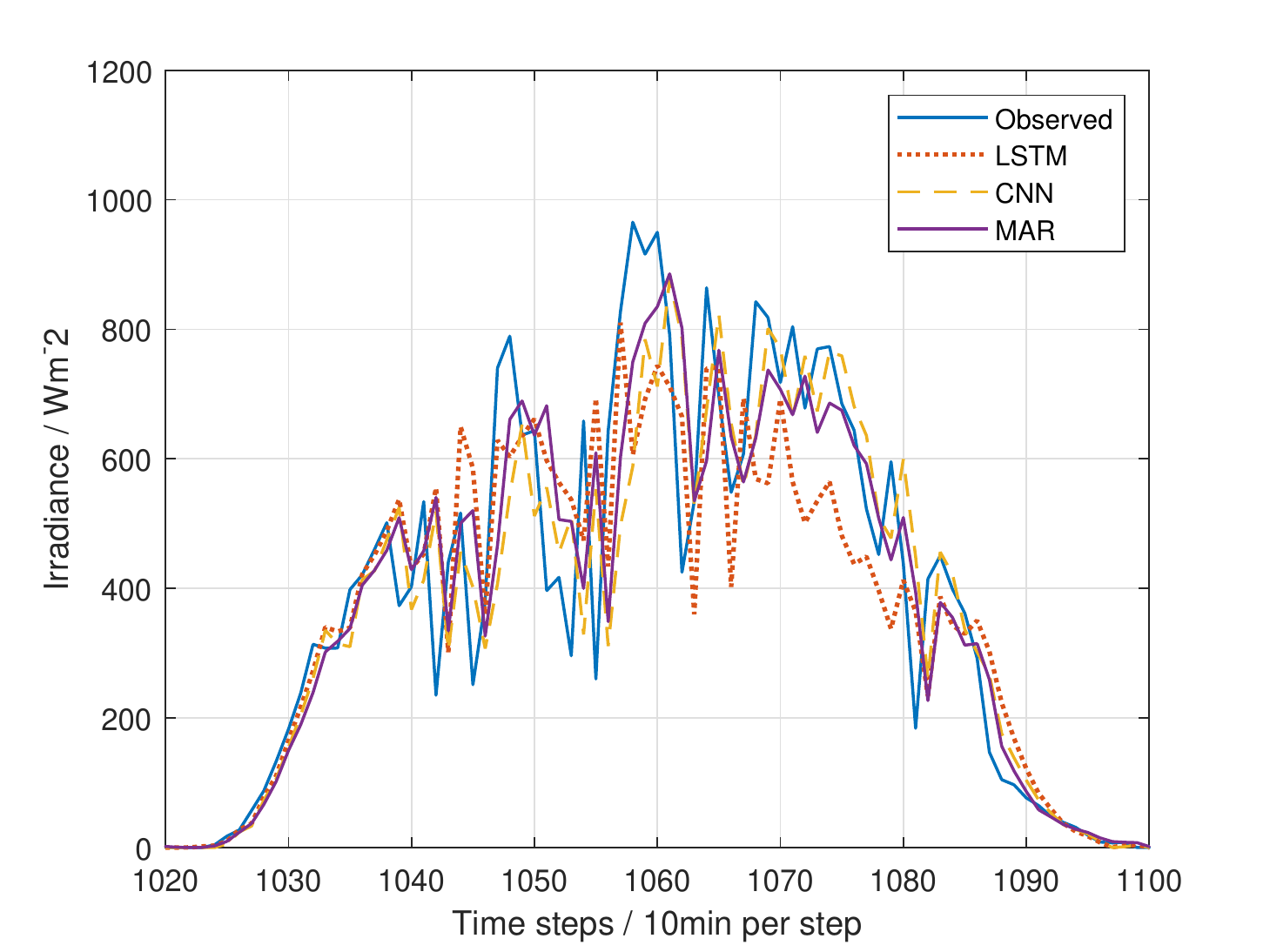}
    \caption{Observed vs. Predicted irradiance for 10 minute prediction horizon extracted for 1 day from the predicted dataset}
    \label{fig:10minpred}
\end{figure}

\begin{figure}[htb]
    \centering
    \includegraphics[width=\columnwidth]{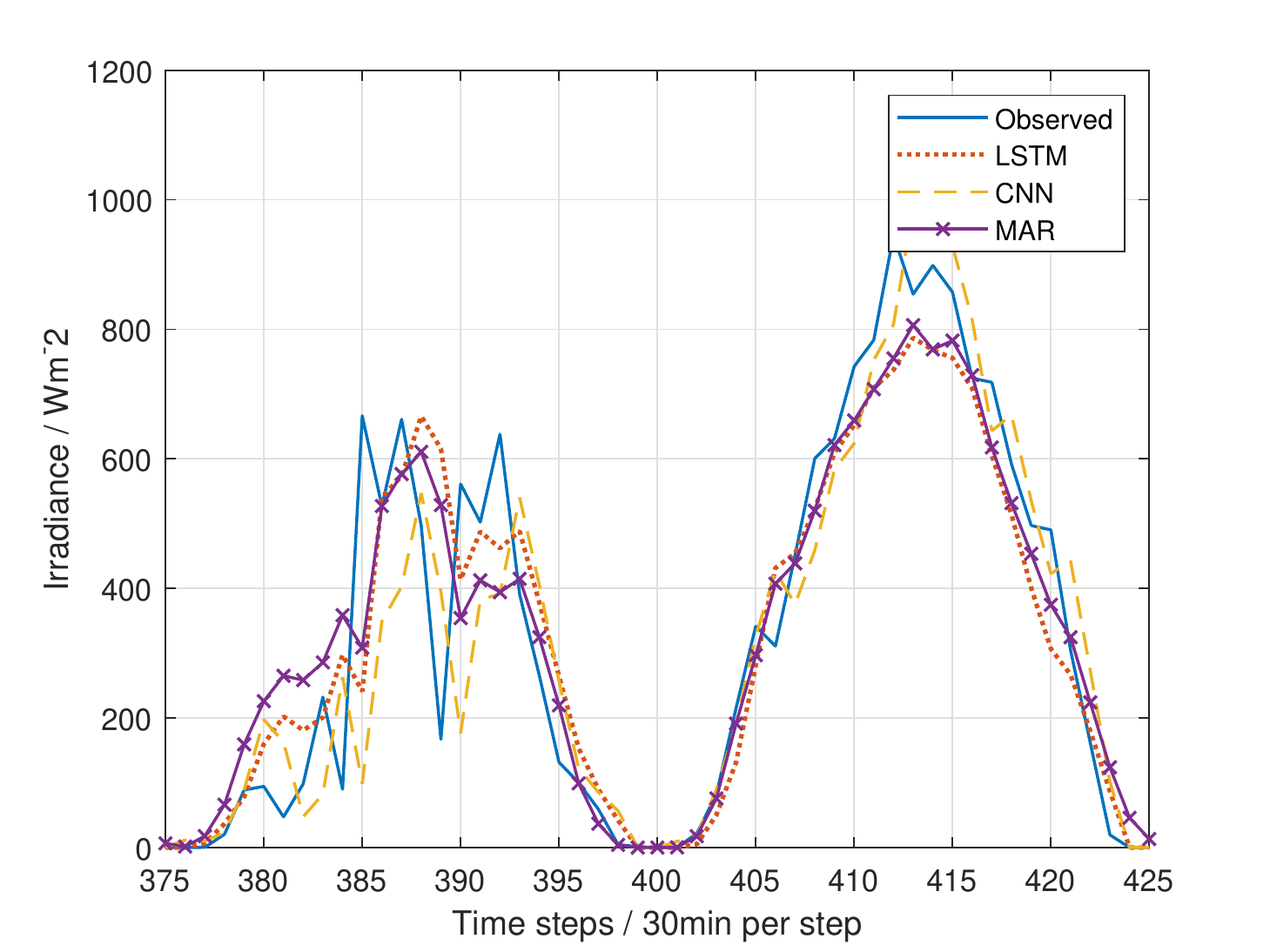}
    \caption{Observed vs. Predicted irradiance for 30 minute prediction horizon extracted for 2 consecutive days from the predicted dataset}
    \label{fig:30minpred}
\end{figure}

\begin{figure}[H]
    \centering
    \includegraphics[width=\columnwidth]{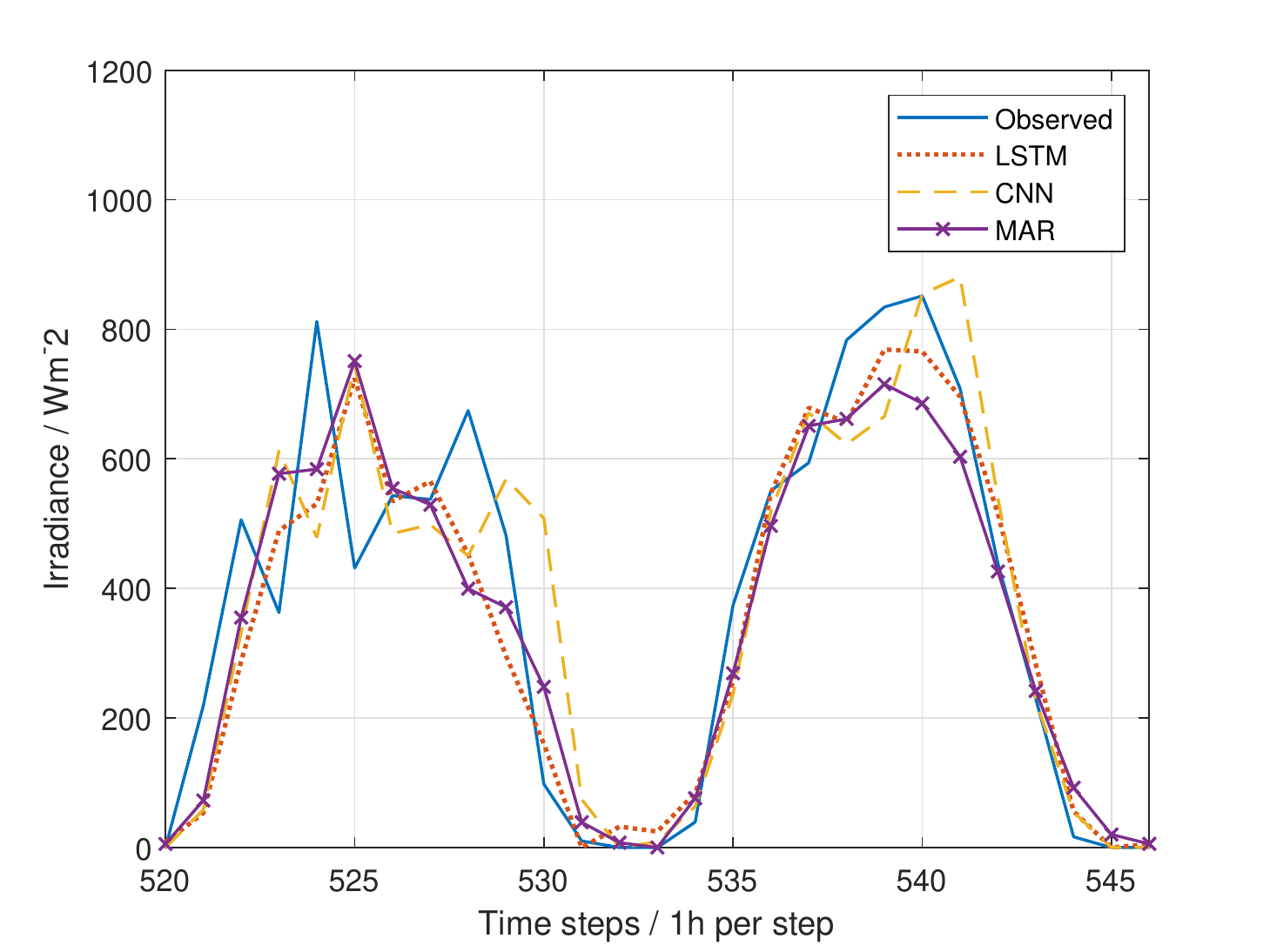}
    \caption{Observed vs. Predicted irradiance for 1 hour prediction horizon extracted for 2 consecutive days from the predicted dataset}
    \label{fig:1hpred}
\end{figure}

Irradiance prediction for two randomly chosen successive days for 10 minute, 30 minute and 1 hour prediction horizons are shown in Fig. \ref{fig:10minpred}, Fig. \ref{fig:30minpred}, Fig. \ref{fig:1hpred} respectively. The deep learning models and the MAR model are designed with one specific model to forecast across all time horizons discussed in the paper. As observed, when the prediction horizon increases, the tendency for predicted curves to follow sudden changes in irradiance is less. Therefore, the dependency on the ensemble (bell-shaped) features of data increases.\par

\begin{table}[htb]
\caption{Error comparison of Prediction Models}
\label{table:Errors}
\begin{center}
\begin{tabular}[\columnwidth]{ l l r r r r }

    \toprule
    \textbf{Error Model} & \textbf{Horizon}   & \textbf{CNN} & \textbf{AR}  & \textbf{LSTM}  & \textbf{MAR}\\
    \midrule
                    &10 min	    &113.62     &115.62     &114.79     &110.38\\
    RMSE /$Wm^{-2}$ &30 min	    &164.63     &170.05     &146.50     &148.25\\
                    &1 h	    &181.98     &182.17     &161.40     &158.56\\
    \midrule
    	            &10 min	    &64.66      &74.52      &69.70      &68.21\\
    MAE /$Wm^{-2}$   &30 min	    &102.82     &123.16     &98.34      &99.06\\
                    &1 h	    &124.52     &138.07     &111.98     &112.09\\
    \midrule
     	            &10 min	    &14.56      &16.02      &14.71      &14.20\\
    MAPE /\%       &30 min	    &21.81      &23.67      &19.36      &19.94\\
                    &1 h	    &24.59      &27.45      &22.18      &22.42   \\
   
    \bottomrule
\end{tabular}
\end{center}
\end{table}

The performance of the four models are evaluated with respect to the metrics of measure Root Mean Square Error (RMSE), Mean Absolute Error (MAE) and Mean Absolute Percentage Error (MAPE) in Table \ref{table:Errors}. It can be observed that the error increases for all models when the prediction horizon increases. However, the performance of the CNN and conventional AR model deteriorates faster than the other two. It is noteworthy that, the MAR model, being a simplistic implementation with pre processing, consolidates a robust performance with the time horizon change while matching the performance of a deep learning LSTM model in all aspects; both errors and increased time horizons.

\section{Conclusion}
\label{sec:conclusion}

In this paper we propose three models of solar prediction; a Modified Auto Regressive (MAR) model, two deep learning models each based on CNN and LSTM neural networks. The performance of the models are quantified by the error metrics RMSE, MAE and MAPE, and it affirms that the MAR model fits best for the case of very-short term prediction of solar irradiance.\par

In a system such as a tropical environment, variability of irradiance at a given timestamp is high, reducing the correlation between consecutive samples. Hence, deep neural networks tend to mostly capture the bell-shaped nature of solar irradiance, as intra-day variations are highly uncorrelated. By means of the ensemble mean curve deduction the MAR, having the least computational cost, is capable of predicting solar irradiance with a performance similar to LSTM- the state of the art prediction scheme- across all tested prediction horizons.\par
Existing prediction models use multi-sensory data; such as temperature, humidity, cloud cover and irradiance. The proposed MAR uses a single sensor measurement as input for the prediction sufficing in performance for most use cases, with an MAPE of less than 15\% for 10 minute prediction, and less than 20\% for 30 minute prediction. This enables an easy acquisition of data, which facilitates an easily deployable 

\noindent forecast system. Thus, taking into account the aforementioned conditions, MAR is chosen as the optimal solar irradiance prediction model. \par

\bibliographystyle{IEEEtran}
\bibliography{IEEEabrv, references}

\end{document}